\documentclass[12pt,journal,onecolumn,draftclsnofoot]{IEEEtran}
\usepackage{amsmath}
\usepackage{graphicx}
\usepackage{amssymb}
\usepackage{amsfonts}
\usepackage{amsthm}
\usepackage{cite}
\usepackage{bm,comment}
\usepackage{algorithm}
\usepackage{algpseudocode}

\usepackage{subeqnarray}
\usepackage{subfigure}
\usepackage{multirow}
\usepackage{stfloats}
\usepackage{float}
\usepackage{color} 
\usepackage{cases}

\newtheorem{proposition}{Proposition}

\floatname{algorithm}{Algorithm}

\begin{document}
\IEEEoverridecommandlockouts
\title{Beamforming Optimization for Intelligent Reflecting Surface Assisted MIMO: A Sum-Path-Gain Maximization Approach}
\author{Boyu Ning, Zhi Chen, \IEEEmembership{Senior Member, IEEE},\\ Wenjie Chen, and Jun Fang, \IEEEmembership{Senior Member, IEEE}\vspace{-6pt}
\thanks{This work was supported in part by the National Key R$\&$D Program of China under Grant 2018YFB1801500.}
\thanks{B. Ning, Z. Chen, W. Chen, and Jun Fang are with National Key Laboratory of Science and Technology on Communications, University of Electronic Science and Technology of China, Chengdu 611731, China (e-mails: boydning@outlook.com; chenzhi@uestc.edu.cn; wenjiechenuestc@163.com; junfang@uestc.edu.cn).}}
\maketitle

\begin{abstract}
Recently, intelligent reflecting surface (IRS) has emerged as an appealing technique that enables wireless communications with low hardware cost and low power consumption. In this letter, we consider an IRS-assisted point-to-point multi-input multi-output (MIMO) system, where a source communicates with its destination with the help of an IRS. Our goal is to maximize the spectral efficiency of this system by jointly optimizing the (active) precoding at the source and the (passive) phase shifters (PSs) at the IRS. However, this turns out to be an intractable mixed-integer non-convex optimization problem. To circumvent the intractability, we propose a new sum-path-gain maximization (SPGM) criterion to obtain a high-quality and efficient suboptimal solution to this problem. Specifically, the PSs are first designed based on a simplified optimization problem, which aims to maximize the sum-gains of the spatial paths between the source and the destination. Then, a low-complexity alternating direction method of multipliers (ADMM) algorithm is utilized to solve this simplified problem. Finally, with the above obtained PSs, the source precoding is derived by performing the singular value decomposition (SVD) on the effective channel between the source and the destination. Numerical results demonstrate that the proposed scheme can achieve near-optimal performance.
\end{abstract}
\IEEEpeerreviewmaketitle


\section{Introduction}
Traditional cooperative transmission technologies, such as amplify-and-forward and decode-and-forward relaying\cite{yfan}, have been widely adopted in wireless communications to improve the network spectrum efficiency by coordinating among distributed equipments. However, the required hardware complexity and system cost are considered as the main hindrance to their large-scale practical implementation\cite{SZhang}. Recently, intelligent reflecting surface (IRS) has emerged as an innovative and cost-effective paradigm for improving the network spectrum efficiency via \emph{passive} reflecting arrays\cite{cometa}. Unlike conventional passive reflectors which retain fixed phase shifters (PSs) once fabricated, IRS is able to induce a certain phase shift independently with the incident electromagnetic beam via controllable meta-material\cite{fujia}. Compared to traditional relaying schemes that enhance source-destination transmission by generating new signals, IRS does not buffer or process any incoming signals but only reflects the wireless signal as a passive planar array, thus incurring no additional power consumption. 

Motivated by the above, many existing works have looked into the spectrum efficient transmit beamforming and IRS PS design in IRS-assisted multi-input single-output (MISO) communication\cite{qinte,qinte2,huangchi,ghy,huangchi2}. It was shown that besides the transmit beamforming gain as in conventional MISO communication, IRS further provides a new aperture gain in enhancing the network spectrum efficiency. Specifically, by properly adjusting the PS at each IRS element, the signal reflected by IRS and that propagated through the direct link can be constructively combined at the receiver, thereby significantly enhancing its received signal power. Nevertheless, due to the more challenging system setup, there has been very limited work\cite{cp} on IRS-assisted multi-input multi-output (MIMO) communication. In \cite{cp}, a sophisticated block coordinate descent (BCD) algorithm was proposed to maximize the weighted sum-rate of an IRS-assisted multi-cell multi-user MIMO network.

In this letter, we propose a more efficient algorithm than \cite{cp} tailored to the IRS-assisted point-to-point MIMO communication between a source node and a destination node, named Alice and Bob, respectively. Our goal is to maximize the system spectral efficiency by jointly optimizing the (active) precoding at Alice and the (passive) PSs at the IRS. Due to the difficulty in directly solving this problem, we first propose a new sum-path-gain maximization (SPGM) criterion and accordingly formulate a simplified problem, which aims to maximize the sum-gains of the spatial paths between Alice and Bob. Then, an efficient PS solution is derived by solving this problem using a low-complexity alternating direction method of multipliers (ADMM) algorithm. With this PS solution, Alice's precoding matrix is obtained by performing the singular value decomposition (SVD) on its effective channel with Bob via IRS. 
\begin{figure}[t]
\centering
\includegraphics[width=3.5in]{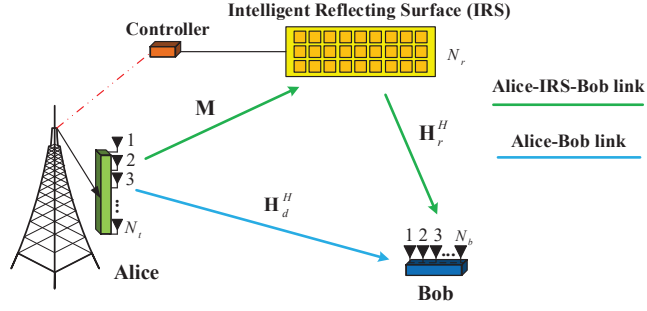}
\caption{An IRS-assisted point-to-point MIMO communication system.}\label{MISOIRS}
\vspace{-12pt}
\end{figure}


\section{System Model and Problem Formulation}
Consider a point-to-point MIMO communication system as depicted in Fig.\,\ref{MISOIRS}, where Alice, equipped with $N_t$ antennas, transmits $N_s \le N_t$ data streams to Bob, equipped with $N_b$ antennas, with the help of an IRS equipped with $N_r$ passive elements and installed on a surrounding wall. We assume that the global channel state information (CSI) is available at a centralized controller, e.g., by applying the cascaded channel estimation method proposed in\cite{hzq}. In the communication, Alice sends its data message ${\bf{s}} \in {\mathbb{C}^{{N_s} \times 1}}, {\bf{s}} \sim \mathcal{CN}({\bf{0}},{\bf{I}}_{N_s})$ via a linear precoder ${\bf{F}} \in {\mathbb{C}^{{N_t} \times {N_s}}}$ to Bob and the IRS simultaneously. Let ${\bf{H}}_d^H \in {\mathbb{C}^{{N_b} \times {N_t}}}$, ${\bf{M}} \in {\mathbb{C}^{{N_r} \times {N_t}}}$, and ${\bf{H}}_r^H \in {\mathbb{C}^{{N_b} \times {N_r}}}$ denote the channels from Alice to Bob, from Alice to the IRS, and from the IRS to Bob, respectively. The received signal at the IRS is first phase-shifted by a diagonal reflection matrix ${\bf{\Theta }} = {\rm{diag}}(\beta {e^{j{\theta _1}}},\beta {e^{j{\theta _2}}}, \cdots ,\beta {e^{j{\theta _{N_r}}}}) \in {\mathbb{C}^{N_r \times N_r}}$ and then reflected to Bob, where $j = \sqrt {-1}$ is an imaginary unit, $\theta _i \in [0,2\pi), i=1,2,\cdots,N_r$ are shifted phases, and $\beta \in [0,1]$ denote the amplitude of each reflection coefficient. Here, we ignore the signals that are reflected by the IRS two or more times. As such, the overall received signal comprising the direct Alice-Bob link and the cascaded Alice-IRS-Bob link is expressed as\cite{qinte2}
\begin{equation}
{{\bf{y}}_b} = \sqrt {\frac{P}{{{N_s}}}} ( {\underbrace {{\bf{H}}_r^H{\bf{\Theta MFs}}}_{{\text{Alice-IRS-Bob}}} + \underbrace {{\bf{H}}_d^H{\bf{Fs}}}_{{\text{Alice-Bob}}}} ) + {{\bf{n}}_b},
\end{equation}
where $P$ is the total transmitted power and ${\|{\bf{F}}\|_F^2} = N_s$. In addition, ${\bf{n}}_b \sim \mathcal{CN}({\bf{0}},\sigma _n^2{{\bf{I}}_{{N_b}}})$ is zero-mean additive Gaussian noise. This letter aims to maximize the spectral efficiency by jointly optimizing the precoding matrix $\bf{F}$ and the PSs $\{\theta _i\}_{i=1}^{N_r}$, subject to the power constraint at Alice and the uni-modular constraints on the PSs. Let ${\bf{v}} = [{e^{j{\theta _1}}},{e^{j{\theta _2}}}, \cdots ,e^{j{\theta _{N_r}}}]^H$ denote the PS vector at the IRS, i.e., ${\bf{\Theta }} = \beta \cdot{\rm{diag}}({\bf{v}}^*)$, with the superscript $(\cdot)^*$ denoting the conjugate operation. Define the effective Alice-Bob channel as ${{\bf{H}}_{\text{eff}}^H} = {\bf{H}}_r^H{\bf{\Theta M}} + {\bf{H}}_d^H$. To maximally exploit the spatial multiplexing gain of MIMO, we consider $N_s = {\rm{rank}}({\bf{H}}_{\text{eff}})$. Our design problem can be formulated as 
\begin{equation}\label{ori}
\begin{split}
&\mathop {\max }\limits_{{\bf{F}},{{\bf{v}}}} \;{\log _2}\det \left| {{{\bf{I}}_{{N_b}}} + \frac{P}{{\sigma _n^2}{N_s}}{{\bf{H}}_{\text{eff}}^H}{\bf{F}}{{\bf{F}}^H}{\bf{H}}_{\text{eff}}} \right|\\
&\;\;\;{\rm{s.t.}}\;\;\;{{\bf{H}}_{\text{eff}}^H} = {\bf{H}}_r^H{\bf{\Theta M}} + {\bf{H}}_d^H,\\
&\;\qquad\;\;{\left\| {\bf{F}} \right\|_F^2} = {N_s},\;{\bf{\Theta }} = \beta \cdot{\rm{diag}}({\bf{v}}^*),\\
&\;\qquad\;\;\left| {{\bf{v}}(i)} \right| = 1, \;\; i=1,2,\cdots,N_r,
\end{split}
\end{equation}
where ${\bf{v}}(i)$ denotes the $i$-th entry of ${\bf{v}}$. As observed from (\ref{ori}), it is difficult to optimally solve (\ref{ori}) due to its non-convex objective function with regard to $\bf v$, as well as the equality constraints therein.

\section{Joint PS and Precoding Design}
In this section, we first propose an efficient SPGM criterion to design the PSs $\bf v$ in (\ref{ori}), and then optimize the precoding matrix $\bf F$ via SVD and water-filling power allocation.

\subsection{PS Design under the SPGM Criterion}
Let ${\bf{H}}_{\rm{eff}}^H = {\bf{U\Lambda }}{{\bf{V}}^H}$ be the truncated SVD of ${\bf{H}}_{\text{eff}}^H$, where ${\bf{U}} \in {\mathbb{C}}^{N_b \times N_s}$ with ${\bf{U}}^H{\bf{U}}={\bf{I}}_{N_s}$, ${{\bf{V}}} \in {\mathbb{C}}^{N_t\times N_s}$ with ${\bf{V}}^H{\bf{V}}={\bf{I}}_{N_s}$, and ${\bf{\Lambda }} \in {\mathbb{C}}^{N_s \times N_s}$ is a strictly positive diagonal matrix with its diagonal elements (singular values) given by $\lambda_1 \ge \lambda_2 \ge \cdots \ge \lambda_{N_s}$. 
Note that for any given ${\bf{H}}_{\rm{eff}}^H$, the optimal precoding matrix should be ${\bf F}_*={\bf{V}}{\bf{\Gamma }}^{\frac{1}{2}}$, where ${\bf{\Gamma }} \triangleq {\rm{diag}}([p_1,p_2,\cdots,p_{N_s}])$ is an $N_s \times N_s$ diagonal matrix with $p_i \ge 0$ denoting the transmit power allocated to the $i$-th data stream, $i=1,2,\cdots,N_s$ and $\sum\nolimits_{i = 1}^{N_s}{p_i}=N_s$. By this means, the effective channel ${\bf{H}}_{\text{eff}}^H$ can be decoupled into $N_s$ parallel single-input single-output (SISO) spatial paths, with $\lambda_i^2$ being the gain of the $i$-th path, $i=1,2,\cdots,N_s$. Accordingly, the spectral efficiency can be re-expressed as $\sum\nolimits_{i = 1}^{N_s}{\log _2}(1+\frac{Pp_i\lambda_i^2}{\sigma_n^2N_s})$. Problem (\ref{ori}) is thus equivalent to
\begin{equation}\label{MIMO}
\begin{split}
\tilde R=\mathop {\max }\limits_{\{p_i\},\bf{v}} &\;\sum\limits_{i = 1}^{N_s}{\log _2}(1+\frac{Pp_i\lambda_i^2}{\sigma_n^2N_s})\\
\;\;\;{\rm{s.t.}}\;&\sum\limits_{i = 1}^{N_s}{p_i}=N_s, \\
&\left| {{\bf{v}}(i)} \right| = 1, \;\; i=1,2,\cdots,N_r,
\end{split}
\end{equation}
where $\{\lambda_i\}_{i=1}^{N_s}$ are dependent on $\bf{v}$ via the SVD of ${\bf{H}}_{\text{eff}}^H$. For problem (\ref{MIMO}), the optimal power allocations $\{p_i\}_{i=1}^{N_s}$ can be easily determined by the water-filling procedure with any given ${\bf{v}}$ (and thus $\{\lambda_i\}_{i=1}^{N_s}$). Thus, we focus on the design of the PSs ${\bf{v}}$ next. 

Unfortunately, it is still difficult to find the optimal ${\bf{v}}$ due to the implicit relation between $\{\lambda_i\}_{i=1}^{N_s}$ and $\bf{v}$, as well as the uni-modular constraints on $\bf{v}$. Nevertheless, it is observed from (\ref{MIMO}) that $\{\lambda_i\}_{i=1}^{N_s}$ indicate the quality of the effective channel ${\bf{H}}_{\text{eff}}^H$. Obviously, the optimal value of (\ref{MIMO}) increases monotonically with $\{\lambda_i\}_{i=1}^{N_s}$. This thus offers us an efficient and reasonable SPGM criterion to solve (\ref{MIMO}) suboptimally, given the obstacle to solving it optimally, as stated below.

\emph{{\bf{SPGM criterion}}: To enhance the channel quality of ${\bf{H}}_{\text{eff}}^H$, we propose to maximize the sum-gain of the spatial paths between Alice and Bob, i.e., $\sum\nolimits_{i = 1}^{{N_s}} \lambda_i^2$.}

Since $\lambda_i^2$ is the $i$-th eigenvalue of ${\bf{H}}_{\text{eff}}^H{\bf{H}}_{\text{eff}}, i=1,2,\cdots,N_s$, we have $\sum\nolimits_{i = 1}^{{N_s}} {\lambda_i^2}={\rm{Tr}}({\bf{H}}_{\text{eff}}^H{\bf{H}}_{\text{eff}})$ and
\begin{align}
\arg \mathop {\max }\limits_{\bf{v}} \sum\limits_{i = 1}^{{N_s}} {\lambda_i^2} &= \arg \mathop {\max }\limits_{\bf{v}} {\rm{Tr}}({\bf{H}}_{\text{eff}}^H{\bf{H}}_{\text{eff}})\\
&= \arg \mathop {\max }\limits_{\bf{v}} {\log _2}[1+{\rm{Tr}}(\frac{P}{{\sigma _n^2}{N_s}}{\bf{H}}_{\text{eff}}^H{\bf{H}}_{\text{eff}})],\nonumber
\end{align}
Thus, the PS optimization problem under the SPGM criterion can be formulated as
\begin{equation}\label{msg}
\begin{split}
\hat R=\mathop {\max }\limits_{\bf{v}} &\;{\log _2}[1+{\rm{Tr(}}{} \frac{P}{{\sigma _n^2}{N_s}}{\bf{H}}_{\text{eff}}^H{\bf{H}}_{\text{eff}}{\rm{)}}]\\
\;\;\;{\rm{s.t.}}\;&{{\bf{H}}_{\text{eff}}^H} = {\bf{H}}_r^H{\bf{\Theta M}} + {\bf{H}}_d^H, \;{\bf{\Theta }} = \beta \cdot{\rm{diag}}({\bf{v}}^*)\\
\;\qquad\;&\left| {{\bf{v}}(i)} \right| = 1, \;\; i=1,2,...,N_r.
\end{split}
\end{equation}
It is worth mentioning that the new problem (\ref{msg}) is closely related to (\ref{MIMO}) via the following two propositions.
\begin{proposition}
The optimal value of (\ref{msg}) is a lower bound on that of (\ref{MIMO}), i.e., \[\hat R \le \tilde R.\] Moreover, in the case of MISO, we have $\hat R = \tilde R$.
\end{proposition}
\begin{IEEEproof}
Assume that the optimal power allocation in (\ref{MIMO}) is equal power allocation, i.e., $p_i=1, i=1,2,\cdots,N_s$. Evidently, we have 
\[\begin{split}
2^{\tilde R} &\ge \prod\limits_{i = 1}^{N_s} {\left(1 + \frac{P\lambda _i^2}{\sigma^2_nN_s}\right)}  \\
&= 1 + \frac{P}{{\sigma _n^2}{N_s}}{\rm{Tr}}({\bf{H}}_{\text{eff}}^H{\bf{H}}_{\text{eff}}) + \sum\limits_{i \ne k} {\frac{P^2}{{\sigma _n^4}{N_s^2}}{\lambda^2_i}{\lambda^2_k}}  + ...\ge 2^{\hat R},\\
\end{split}\]
It is easy to see that in the case of MISO, i.e., $N_s=N_b=1$, we have $\tilde R = \hat R$.
\end{IEEEproof}

\begin{proposition}
The optimal value of (\ref{msg}) provides the following upper bound on that of (\ref{MIMO}),
\[\tilde R \le N_s{\hat R}.\]
\end{proposition}
\begin{IEEEproof}
Note that 
\begin{equation}\label{eq1}
{2^{\tilde R}} = \prod\limits_{i = 1}^{N_s} {(1 + \frac{Pp_i\lambda_i^2}{{\sigma _n^2N_s}})} \mathop  \le \limits^{(a)} \frac{1}{{N_s}^{N_s}}{\left[ {N_s + \frac{P}{{\sigma _n^2N_s}}\sum\limits_{i = 1}^{N_s} {p_i\lambda_i^2} } \right]^{N_s}},
\end{equation}
where equality ($a$) is due to the inequalities of arithmetic and geometric means\cite{Golub1996}.  

On the other hand, it must hold that
\[\sum\nolimits_{i = 1}^{N_s} {p_i\lambda_i^2} \le \sum\nolimits_{i = 1}^{N_s} {p_i}\sum\nolimits_{j = 1}^{N_s} {\lambda_j^2} = N_s\sum\nolimits_{j = 1}^{N_s} {\lambda_j^2}.\]
By substituting this into (\ref{eq1}), we obtain
\[{2^{\tilde R}} \le \frac{1}{{N_s}^{N_s}}{\left[ {N_s \!+\! \frac{P}{{\sigma _n^2}}\sum\limits_{i = 1}^{N_s} {\lambda_i^2} } \right]^{N_s}}\!\!\!\!=\!\!{\frac{1}{{N_s}^{N_s}}{{\left[ {N_s \!+\! N_s({2^{{\hat R}}} - 1)} \right]}^{N_s}}}.\]

Next, we take the binary logarithm at both sides of the above inequality and obtain
\begin{equation}\label{pop2}
\tilde R \le {\log _2}\left\{ {\frac{1}{{N_s}^{N_s}}{{\left[ {N_s + N_s({2^{{\hat R}}} - 1)} \right]}^{N_s}}} \right\}=N_s{\hat R}.
\end{equation}
Proposition 2 is thus proved.
\end{IEEEproof}
Although the bounds ${\hat R} \le \tilde R \le N_s{\hat R}$ become looser as $N_s$ increases, the proposed SPGM criterion is effective in the case of small and moderate $N_s$, as will be shown in Section \ref{nu}. To solve the SPGM-based problem (\ref{msg}), let us re-express (\ref{msg}) in a more explicit equivalent form:
\begin{equation}\label{ms}
\begin{split}
&\mathop {\max }\limits_{\bf{v}}\;\;{\rm{Tr}}\big[({\bf{H}}_r^H{\bf{\Theta M}} + {\bf{H}}_d^H){({\bf{H}}_r^H{\bf{\Theta M}} + {\bf{H}}_d^H)^H}\big]\\
&\quad{\rm{s.t.}}\quad {\bf{\Theta }} = \beta \cdot{\rm{diag}}({{\bf{v}}^*}),\;\left| {{\bf{v}}(i)} \right| = 1,\;i = 1,...,{N_r}.
\end{split}
\end{equation} 
The objective function of (\ref{ms}) can be equivalently rewritten as $\sum\nolimits_{i=1}^{{N_b}} {\left[ {{\bf{\Pi }}(i,i) + {\bf{\Psi }}(i,i) + {{\bf{\Psi }}^H}(i,i)} \right]}  + C$, where ${\bf{\Pi }} = {\bf{H}}_r^H{\bf{\Theta M}}{{\bf{M}}^H}{{\bf{\Theta }}^H}{{\bf{H}}_r}$, ${\bf{\Psi }} = {\bf{H}}_r^H{\bf{\Theta M}}{{\bf{H}}_d}$, and $C = {\rm{Tr(}}{\bf{H}}_d^H{{\bf{H}}_d}{\rm{)}}$ is independent of $\bf{v}$. Let ${{\bf{H}}_{r}}= \left[ {{{\bf{h}}_1},{{\bf{h}}_2},...,{{\bf{h}}_{{N_b}}}} \right]$ and ${\bf{M}}{{\bf{H}}_d} = [{{\bf{k}}_1},{{\bf{k}}_2},...,{{\bf{k}}_{{N_b}}}]$. Then, it can be verified that
\begin{equation}
\begin{split}
{\bf{\Pi }}(i,i) &= {\beta ^2}{{\bf{h}}_i^H{\rm{diag}}({{\bf{v}}^*}){\bf{M}}{{\bf{M}}^H}{\rm{diag}}({\bf{v}}){{\bf{h}}_i}}, \;\\
{\bf{\Psi }}(i,i) &= \beta  {{\bf{h}}_i^H{\rm{diag}}({{\bf{v}}^*}){{\bf{k}}_i}}. 
\end{split}
\end{equation}
By applying the fact that ${\bf{h}}_k^H{\rm{diag}}({{\bf{v}}^*}) = {{\bf{v}}^H}{\rm{diag}}({\bf{h}}_k^H)$, we tactfully introduce the following correlative matrix and vector,
\begin{equation}\label{Tmatrix}
{\bf{T}} = \left[ {\begin{array}{*{20}{c}}
-{{\beta ^2}\sum\limits_{i = 1}^{{N_b}} {{\rm{diag}}({\bf{h}}_i^H){\bf{M}}{{\bf{M}}^H}{\rm{diag}}({{\bf{h}}_i})}}\!\!&\!\!{-\beta \sum\limits_{i = 1}^{{N_b}} {{\rm{diag}}({\bf{h}}_i^H){{\bf{k}}_i}} }\\
-{\beta \sum\limits_{i = 1}^{{N_b}} {{\bf{k}}_i^H{\rm{diag}}({{\bf{h}}_i})} }\!\!&\!\!0
\end{array}} \right]
\end{equation}
and ${\bf{x}} = {[t \cdot {{\bf{v}}^H},\;\;t]^H}$, respectively, where $t \in \mathbb{C}$ is an auxiliary variable satisfying $\lvert t \rvert = 1$. Thus, (\ref{ms}) can be equivalently rewritten in the following form amenable to ADMM\cite{liqiang}, i.e.,\vspace{-9pt}
\begin{subequations}\label{admm}
\begin{align}
&\mathop {\min }\limits_{{\bf{x}},{\bf{u}} \in \mathbb{C}^{N_r + 1}} \;\;\frac{1}{2}{{\bf{x}}^H}\widehat {\bf{T}}{\bf{x}}\\
&\;{\rm{s.t.}}\;\;\lvert {\bf{u}}(i)\rvert=1\;\;i = 1,2,\cdots,{N_r+1},\; {\bf{u}} = {\bf{x}},\label{adc}
\end{align}
\end{subequations}
where $\widehat {\bf{T}} \triangleq {\bf{T}} - {\lambda _{\rm{min}}}({\bf{T}}) \cdot {{\bf{I}}_{N_r+1}} \succeq {\bf{0}}$. Here, $\lambda _{\rm{min}}({\bf{T}})$ denotes the smallest eigenvalue of ${\bf{T}}$. For problem (\ref{admm})\footnote{Note that problem  (\ref{admm})  can also be solved by the semi-definite relaxation (SDR) technique \cite{qinte}. As will be shown in Section IV, both algorithms achieve a comparable performance. However, the SDR technique induces much higher computational complexity of $O(N_r^6)$, as compared to our proposed ADMM algorithm with the complexity of $O(N_r^3 )$.}, its augmented Lagrangian function is given by
\begin{equation}
{\cal L}({\bf{x}},{\bf{u}},{\bm{\nu }}) = \frac{1}{2}{{\bf{x}}^H}\widehat {\bf{T}}{\bf{x}} + {\mathop{\rm Re}\nolimits} \left\{ {{{\bm{\nu }}^H}(\;{\bf{u}} - {\bf{x}})} \right\} + \frac{\rho }{2}{\left\| {{\bf{u}} - {\bf{x}}} \right\|^2},
\end{equation}
where ${\bm{\nu }} \in \mathbb{C}^{N_r + 1} $ is the Lagrange multiplier corresponding to the constraint (\ref{adc}) and the penalty parameter $\rho $ is positive. Let $({{\bf{u}}^0},{{\bf{x}}^0},{{\bm{\nu }}^0})$ be the initial primal-dual variables. The standard ADMM consists of the following iterative procedures:
 \begin{numcases}{}
{{\bf{u}}^{k + 1}} = \arg \mathop {\min }\limits_{{\bf{u}}, \lvert{\bf{u}}(i)\rvert=1\hfill} {\cal L}({\bf{u}},{{\bf{x}}^k},{{\bm{\nu }}^k}).\label{ad1}\\
{{\bf{x}}^{k + 1}} = \arg \mathop {\min }\limits_{\bf{x}} {\cal L}({{\bf{u}}^{k + 1}},{\bf{x}},{{\bm{\nu }}^k}).\label{ad2}\\
{{\bm{\nu }}^{k + 1}} = {{\bm{\nu }}^k} + \rho ({{\bf{u}}^{k + 1}} - {{\bf{x}}^{k + 1}}).\label{ad3}
\end{numcases}
It can be easily verified that problem (\ref{ad1}) admits a closed-form solution ${{\bf{u}}^{k + 1}} = \angle ({{\bf{x}}^k} - {\rho ^{ - 1}}{{\bm{\nu }}^k})$, where $\angle (\bf{\cdot})$ represents the phase vector of its argument. The subproblem (\ref{ad2}) is an unconstrained least-squares problem. By taking the first-order derivative and setting it to zero, we get
\begin{equation}\label{yijie}
\widehat {\bf{T}}{{\bf{x}}^{k + 1}} - {{\bm{\nu }}^k} - \rho ({{\bf{u}}^{k + 1}} - {{\bf{x}}^{k + 1}}) = 0.
\end{equation}
Rearranging (\ref{yijie}) yields 
\begin{equation}\label{bfu}
{{\bf{x}}^{k + 1}}{\rm{ = (}}\rho {\bf{I}} + \widehat {\bf{T}}{)^{ - 1}}(\rho {{\bf{u}}^{k + 1}} + {{\bm{\nu }}^k}).
\end{equation}
Moreover, by following (\ref{ad3}) and (\ref{yijie}), we have 
\begin{equation}\label{bfnu}
{{\bm{\nu }}^{k + 1}} = \widehat {\bf{T}}{{\bf{x}}^{k + 1}}.
\end{equation}
Here, it is worth noting that the proposed ADMM process for solving (\ref{admm}) has a convergence guarantee if the penalty parameter satisfies $\rho  \ge \max \{ \sqrt {2{\lambda _{\rm{max}}}(\widehat {\bf{T}})} ,{\lambda _{\rm{max}}}(\widehat {\bf{T}})\}$ (see Proposition 1 in \cite{liqiang}), where $\lambda _{\rm{max}}$ denotes the largest eigenvalue. Finally, the latent solution $\bf{v}$ to problem (\ref{ms}) can be recovered as
\begin{equation}
{\bf{v}} = {\left\{ {\frac{{{{\bf{x}}^k}}}{t}} \right\}_{(1:{N_r})}} = {\left\{ {{{\left[ {{{\bf{x}}^k}({N_r} + 1)} \right]}^{ - 1}}{{\bf{x}}^k}} \right\}_{(1:{N_r})}},
\end{equation}
where ${\bf{a}}_{(1:N)}$ denotes the vector that contains the first $N$ elements of $\bf{a}$.\vspace{-8pt}

\subsection{Precoding Design with the Obtained PSs}\label{precoding}
Next, we optimize the precoding matrix $\bf F$ (or the power allocation $\{p_i\}_{i=1}^{N_s}$ in (\ref{MIMO})) with the PSs obtained in the last subsection. First, the optimal power allocation in (\ref{MIMO}) can be found by employing the water-filling procedure\cite{boyd}, i.e.,
$
p_i^*=\left(\frac{1}{\rho\ln 2}-\frac{\sigma_n^2N_s}{P\lambda_i^2}\right)^+,
$
where $\left(\cdot\right)^+ \triangleq \mathop {\max }\{\cdot,0\}$, and $\rho$ is a constant ensuring $\sum\nolimits_{i = 1}^{N_s}{p_i^*}=N_s$. Let ${\bf \Gamma}_*$ denote the corresponding optimal power allocation matrix. Then, the optimal precoding matrix can be obtained as ${\bf F}_*={\bf{V}}{{\bf{\Gamma }}_*^{\frac{1}{2}}}$. We summarize the proposed approach to solving (\ref{ori}) in Algorithm 1. 
\begin{algorithm}
  \caption{SPGM-based joint PS and precoding design}
  \begin{algorithmic}[1]
  \Require  $P,\; \beta,\; \sigma _n^2,\; {\bf{H}}_d,\;{\bf{M}},\;{{\bf{H}}_{r}}= \left[ {{{\bf{h}}_1},{{\bf{h}}_2},...,{{\bf{h}}_{{N_b}}}} \right]$.
  \State Compute ${\bf{M}}{{\bf{H}}_d} = [{{\bf{k}}_1},{{\bf{k}}_2},...,{{\bf{k}}_{{N_b}}}]$, $\bf T$ in (\ref{Tmatrix}) and $\widehat {\bf{T}} \triangleq {\bf{T}} - {\lambda _{\min}}({\bf{T}}) \cdot {\bf{I}}_{N_r+1}$.
  \State Given initial variables ${{\bf{u}}^0},{{\bf{x}}^0},{{\bm{\nu }}^0}$ s.t. (\ref{adc}), choose $\rho  = \max \{ \sqrt {2{\lambda _{\rm{max}}}(\widehat {\bf{T}})} ,{\lambda _{\rm{max}}}(\widehat {\bf{T}})\}$. Let $k=0$ and ${R_m}(k)$ denote the objective value of (\ref{admm}) after the $k$th iteration.
  \State \quad \textbf{Repeat}
    \State \qquad ${{\bf{u}}^{k + 1}} \leftarrow \angle ({{\bf{x}}^k} - {\rho ^{ - 1}}{{\bm{\nu }}^k})\;$;
    \State \qquad ${{\bf{x}}^{k + 1}} \leftarrow (\rho {\bf{I}} + \widehat {\bf{T}})^{ - 1} (\rho {{\bf{u}}^{k + 1}} + {{\bm{\nu }}^k})$;
    \State \qquad ${{\bm{\nu }}^{k + 1}} \leftarrow \widehat {\bf{T}}{{\bf{x}}^{k + 1}}$;
    \State \qquad $k \leftarrow k + 1$.
    \State \quad {\textbf{Until}} $\lvert [R_m(k) - R_m(k - 1)]/{R_m}(k - 1) \rvert < \varepsilon$.
 \State Compute ${\bf{v}}^* = {\{ {{{\left[ {{{\bf{x}}^k}({N_r} + 1)} \right]}^{ - 1}}{{\bf{x}}^k}}\}_{(1:{N_r})}}.$
 \State Compute ${{\bf{H}}_{{\text{eff}}}} = \beta {\bf{H}}_r^H{\rm{diag}}({{\bf{v}}^*}){\bf{M}} + {\bf{H}}_d^H$ and determine the optimal precoding matrix ${\bf{F}}^*$ via the procedures given in Section \ref{precoding}.
  \Ensure   ${\bf{v}}^*$ and ${\bf{F}}^*$.
    \end{algorithmic}
\end{algorithm}

\section{Numerical Results}\label{nu}
In this section, numerical results are provided to demonstrate the performance of Algorithm 1 as compared to different benchmarks. We assume that Alice, IRS, and Bob are located at the vertices of an equilateral triangle whose length of each side is $d$ in meter (m). We consider the Rician fading channel model for all channels involved\cite{qinte}, i.e.,
\begin{equation}\label{model}
{\bf{H}} = {\sqrt{L(d)}}\left(\sqrt {\frac{{\kappa }}{{1+\kappa}}} {{\bf{a}}_r}(\phi_r^{}){{\bf{a}}_t}{(\phi _t^{})^H} + \sqrt {\frac{1}{{1 \!+\! {\kappa }}}} {\bf{H}}^{{\rm{NLoS}}}\right)
\end{equation}
for ${\bf{H}}_d^H$, ${\bf{H}}_r^H$, and ${\bf{M}}$. In (\ref{model}), $L(d) = {C_0}{(d/{D_0})^{ - a}}$ is the distance-dependent path-loss factor, where $C_0$ is the path-loss at the reference distance $D_0=1$ m, and $a$ denotes the path-loss exponent. $\kappa$  is the Rician factor and the variables $\phi _{t}$ and $\phi_r \in [0,2\pi )$ are the azimuth angles of departure and arrival for the line-of-sight (LoS) component, respectively. ${{\bf{a}}_t}(\phi )$ and ${{\bf{a}}_r}(\phi )$ are the antenna array response vectors at the transmitter and the receiver, respectively. For simplicity, we consider a uniform linear array configuration with $N$ elements, i.e., ${\bf{a}}(\phi ) = ({1}/{{\sqrt N }}){[1,{e^{jkd_a\sin\phi }},...,{e^{jkd_a(N - 1)\sin\phi }}]^T}$, where $k = 2\pi /\lambda $. $\lambda$ is the wavelength and $d_a$ is the antenna spacing. The entries of ${\bf{H}}_{}^{{\rm{NLoS}}}$ are randomly generated and follow an independent and identically distributed complex Gaussian distribution with zero mean and unit variance. In the simulation, we set $C_0=-30$dB, $d=30$m, $\sigma _n^2 = 1$, $\beta  = 1$, $d_a = \lambda /2$, $\kappa =10$ dB, and $a=2$. The results were averaged over 1,000 random channel realizations.
\begin{figure}[!t]
\centering  
\includegraphics[width=4.8in]{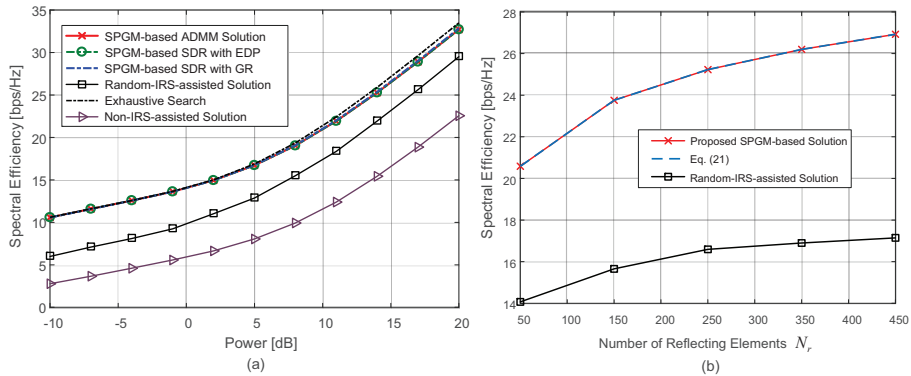}
\caption{(a) Spectral efficiency versus transmit power, with $N_t=N_r=16$ and $N_b=4$. (b) Spectral efficiency versus IRS elements number $N_r$, with $N_t=16$, $N_b=4$, and $P=10$ dB.}
\vspace{-13pt}
\end{figure}

First, Fig.\,2(a) shows the spectral efficiency versus transmit power achieved by different schemes. Specifically, we compare our proposed SPGM-based ADMM algorithm with the following four benchmarks: 1) SDR-based PS solution to problem (\ref{admm})\footnote{Note that the solution by SDR may not be of rank one. In this case, eigen-decomposition and projection (EDP) onto the constant envelope or Gaussian randomization (GR) is needed to extract a feasible solution.}; 2) randomly generated PSs; 3) exhaustive search over 500,000 random PSs for solving (\ref{ori}), and 4) optimal precoding design without IRS. For benchmarks 1)-3), the precoding design is determined similarly as in Section III-B. It is observed from Fig.\,2(a) that the proposed scheme achieves a comparable performance to the exhaustive search and the SDR-based PS design, yet with considerably lower computational complexity (see footnote 1). Moreover, the proposed scheme is also observed to significantly outperform the schemes with random PS design and  without IRS. This implies that the proposed scheme manages to reap both the beamforming gain and the aperture gain offered by the IRS to improve the Alice-Bob channel quality. To drive insights, Fig.\,2(b) shows the spectral efficiency versus IRS elements number $N_r$ with only the LoS component, i.e., $\kappa=\infty$. It was revealed in \cite{qinte} that in the IRS-assisted MISO system, the signal-to-noise ratio (SNR) at Bob grows quadratically with $N_r$. As observed from Fig.\,2(b), this phenomenon still holds in the IRS-assisted MIMO system, since the spectral efficiency grows at the same rate as the following asymptotic function,
\begin{equation}\label{eq}
{\log _2}\left[ 1+{(P/\sigma _n^2){L^2}(d){N_t}{N_b}N_r^2} \right].
\end{equation}

\section{Conclusions}
We considered the joint design of PS and source precoding in an IRS-assisted point-to-point MIMO system. A SPGM criterion is first proposed to design the PSs at the IRS efficiently, and the corresponding optimal precoding matrix is derived by applying the SVD with water-filling allocations. Numerical results showed that our proposed scheme can achieve near-optimal performance with moderate computational complexity, by exploiting both the conventional beamforming gain and the new aperture gain for IRS.

\end{document}